# A Comment on the Origin of Conductance Dips at Finite Bias in Andreev Reflection Spectroscopy Data

Haibing Peng *

Department of Physics and the Texas Center for Superconductivity, University of Houston,

Houston, Texas 77204-5005, USA

* haibingpeng@uh.edu

## ABSTRACT

In an experimental study of Andreev reflection (AR) in normal metal/superconductor (N-S) devices featuring a superconducting topological insulator $Cu_xBi_2Se_3$ [arXiv:1301.1030], Peng et. al. reported the reproducible experimental observation of archetypical double-peak AR spectra for finite barrier strength in N-S junctions, which clearly indicates the absence of zero-energy peak in density of states and thus rules out theoretically predicted zero-energy Majorana fermions in $Cu_xBi_2Se_3$. This report casts doubt on previous claims of $Cu_xBi_2Se_3$ as a topological superconductor. Here we clarify an incorrect understanding which claims that no spectroscopic information can be extracted from N-S point-contacts showing conductance dips at finite bias. Such a clarification of the origin of conductance dips in AR spectra is important for future research in this field.

## Introduction

In Ref. [1], Peng et. al. have demonstrated that for the same piece of superconductor microcrystal of $Cu_xBi_2Se_3$, a zero-bias conductance peak occurs for N-S junctions with transparent barriers, but is absent in N-S junctions with finite barriers showing archetypical double-peak AR spectra, which rules out the existence of zero-energy surface bound states in $Cu_xBi_2Se_3$. The overall evolution of the AR spectra as a function of magnetic field and temperature can be well described by the result from a single-gap bulk superconductor. Therefore, the zero-bias conductance peak observed in $Cu_xBi_2Se_3$ [Ref. 2] and $Sn_{1-x}In_xTe$ [Ref. 3] does not represent an evidence for the existence of characteristic zero-energy Majorana fermions for topological superconductors.

We note that in experimental AR spectra, conductance dips at finite bias are quite often observed (see, for example, Refs. [1] and [3]), and could not be explained by BTK theory. Here, in the form of questions and answers, we clarify an incorrect understanding which claims that no spectroscopic information can be extracted from N-S point-contacts showing conductance dips at finite bias. The clarification of the origin of conductance dips is important for future research on AR spectroscopy.

## Sec I. The determination of spectroscopic regime in N-S junctions

1. One may comment: "*Before the analysis of the point-contact data it is most important to concentrate on the transport regime of the point-contact. The authors understand this and that is why they have included a short discussion on whether the contacts that they analyzed were in the ballistic regime where spectroscopy could be possible. However, their judgment based on the normal state resistance is erroneous. Simply a high resistance in the normal state does not ensure ballisticity of the point-contacts. There are other recipes to confirm this, but first one should inspect the nature of the spectra carefully. When the sharp conductance-dips are present at finite bias voltages, the spectra are clearly not captured in the ballistic regime. No spectroscopic information can be extracted from such point-contacts. All the spectra that the authors present in this paper have the conductance dips that they cannot fit. Therefore it can be concluded that these experiments do not provide any spectroscopic information about the Majorana Fermions or any other kind of surface states.*"

   We discuss below in details on how to determine the transport regime of a point contact in order to obtain spectroscopic (i.e. energy-resolved) information on the superconducting gap values.

   (i) First, we would like to point out that the use of the normal-state resistance to estimate the point-contact size and thus to determine the ballisticity of a N-S junction is widely adopted as an effective empirical rule in Andreev reflection experiments [please refer to a few review papers on this topic, e.g., page 3158-3159 in *J. Phys.: Condens. Matter* **1** 3157, 1989 (by Duif et al.); page 8 in *arXiv:physics/0312016v1* ( by Naidyuk and Yanson); and page 2-4 in S*upercond. Sci. & Tech.* **23,** 043001, 2010 (by Daghero and Gonnelli)].
Therefore, an assertion that "*their judgment (the determination of transport regime) based on the normal state resistance is erroneous*", is simply unfounded.  We explain the details as follows.

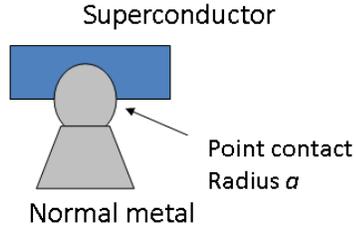

Fig. B1

In principle, the superconducting gap energy can only be measured accurately in either the ballistic transport regime with the actual N-S point contact radius $a$ much less than the electron mean free path $l$, or in the diffusive regime with no significant inelastic scattering (albeit introducing a non-ideal effect of reducing the Andreev refection ratio).  In the so-called thermal regime with significant inelastic scattering, the Andreev reflection spectrum can be distorted by energy-changing scattering events and the gap energy may not be obtained accurately from the experimental data.  For ballistic conduction through a restriction (e.g., a typical point contact with a radius $a$ as shown above in Fig. B1), the electrical resistance at the normal state can be expressed by the Sharvin formula as $R_N = (4\rho l)/(3\pi a^2)$, with $\rho$ the bulk material resistivity of the normal state superconductor.  In the case of only one point contact dominating the conduction in a real N-S junction, the ballistic condition $a \ll l$ can then be turned into a condition on the normal-state contact resistance: $R_N \gg 4\rho/3\pi l$.  This has been commonly used as an empirical criterion for determining ballisticity, since the normal-state resistance $R_N$ can be measured in experiments (but the radius of a real point contact is experimentally inaccessible in general).

In the case of multiple parallel point contacts contributing to the conduction in a N-S junction, the normal-state resistance of each individual contact is larger than the measured total normal-state resistance $R_N$ (i.e. the resistance of the whole N-S junction).  Therefore, as long as the measured normal-state resistance $R_N \gg 4\rho/3\pi l$, the N-S junction (either with single or multiple contacts) showing Andreev reflection should be ballistic.

(ii) Experimentally, if an energy gap value can be reproducibly obtained in different N-S junctions for the same material, one can also conclude that the spectroscopic, energy-resolved condition is met (otherwise, inelastic scattering causes energy change and thus experimental results may vary from sample to sample). Therefore, our reproducible experimental data of similar gap energy (e.g. Fig. 2 of the main text in Ref. [1]) demonstrate clearly a transport regime suitable for obtaining spectroscopic, energy-resolved information.  In addition, we would like to draw attention that the use of the empirical rule to determine the spectroscopic condition based on normal-state-resistance values is well justified by our control experiments (see the Supplemental Material of Ref. [1]).

- (iii) One may further asserts that "*All the spectra that the authors present in this paper have the conductance dips that they cannot fit. Therefore it can be concluded that these experiments do not provide any spectroscopic information about the Majorana Fermions or any other kind of surface states*". We would like to stress that this incorrect statement is based on a lack of understanding on the scientific background related to the conductance dips. Such conductance dips have been often reported in traditional point-contact AR spectra for bulk superconductors, but they are not accounted for by the BTK theory and their physical origin is still under debate [see two different opinions in: Strijkers et al. PHYS. REV. B **63** 104510 (2001); and Sheet et al. PHYS. REV. B **69** 134507 (2004)]. Previously, the proximity effect was proposed [PHYS. REV. B **63** 104510 (2001)] to explain the dips. Alternatively, the role of critical current $I_c$ in the junction was considered and the dips were suspected to be caused by point contacts not being in the ballistic limit [PHYS. REV. B **69** 134507 (2004)]. Recently, we showed via diagnostic experiments that such conductance dips occur when the critical current $I_c$ of the corresponding section of the SC microcrystal is reached, and have nothing to do with the transport regime [see Fig. 3 of J. Phys.: Condens. Matter **24**, 455703 (2012)]. However, no matter what is the origin of the conductance dips, an assertion that "*these (*our*) experiments do not provide any spectroscopic information*" because of the occurrence of the dips, are not accurate if one simply takes a look at the data from the paper by Sheet et al. [e.g., Fig. 2 in Phys. Rev. B **69** 134507 (2004)]. As seen clearly therein, for samples with or without dips, the energy scale of the gap is similar, and thus indeed provide spectroscopic info in contrast to the above cited claim.

2. One may states: "*The authors also need to take care while fitting a spectrum with more than four parameters. Ideally, with four fitting parameters any spectrum could be fitted if all four of them are free. They should note that when the value of Gamma (they found it to be up to 0.27 meV) becomes comparable in magnitude to the superconducting gap (0.35 meV in their case) itself, the situation becomes unphysical.*"

We would like to mention that the current standard practice for fitting AR spectra includes a minimum of three fitting parameters for a one-gap superconductor: the gap value $\Delta$, the barrier strength $Z$, and the broadening parameter $\Gamma$ [See: Plecenik et al., Phys. Rev. B **49**, 10016 (1994), or review papers such as S*upercond. Sci. & Tech.* **23,** 043001, 2010 (by Daghero and Gonnelli)]. In this work, we introduced an additional parameter $w$ besides the other three minimum fitting parameters, in order to account for the weight of contribution to the conductance from the superconducting phase in $Cu_xBi_2Se_3$, which is well justified considering the mixing of superconducting and non-superconducting phases in this material (as commonly reported in literature).

We also cannot agree with the argument that "*when the value of Gamma (they found it to be up to 0.27 meV) becomes comparable in magnitude to the superconducting gap (0.35 meV in their case) itself, the situation becomes unphysical*." The physical origin of the broadening parameter $\Gamma$ includes different sources [see: S*upercond. Sci. & Tech.* **23**, 043001, 2010 and Ref. 36 therein], such as intrinsic quasiparticle lifetime, inelastic scattering near the N-S interface (e.g., due to surface degradation or contaminations), and even a distribution of gap values (e.g. in anisotropic superconductors). In our fitting, $\Gamma$ ranges from 0.13 to 0.27 meV for a gap ~0.35 meV, corresponding to a ratio of $\Gamma/\Delta$ from 0.37 to 0.77. In literatures, similar $\Gamma/\Delta$ ratio has been reported: $\Gamma/\Delta$=0.5-0.7 for Fe-based superconductor [Tortello et al., PHYS. REV. LETT. **105**, 237002 (2010)], and $\Gamma_\pi$ = 2 meV for a gap $\Delta_\pi$ = 2.8 meV in a superconductor $MgB_2$ [Gonnelli et al., PHYS. REV. LETT. **89**, 247004 (2002)]. Therefore, from either theoretical or experimental point of view, it is not justified to conclude that our fitted $\Gamma$ values are "*unphysical*".

## Sec. II. More details on the origin of the conductance dips in AR spectra

To further clarify the origin of the dI/dV dips at finite bias, we reemphasize below the relevant experimental facts and physical explanations already existing in literatures.

1. One may states: "*The authors have cited the work of Sheet et.al. and attempted to prove that the appearance of the conductance dips do not give any information about the transport regime. This is just opposite of what Sheet et.al. themselves claim in the cited publication.*"

The paper of Sheet et al. [Phys. Rev. B **69**, 134507 (2004)] was the first to draw attention to the role of the critical current in the dI/dV dips, but some of their physical interpretation is still under debate (we actually don't fully agree with part of their arguments purely focused on the N-S interface area). Nevertheless, the experimental data which we adapted from the paper of Sheet et. al. themselves (see, e.g., Fig. 2 therein) simply show the fact that the occurrence of the dips does NOT support an argument that "*these (our) experiments do not provide any spectroscopic information*", since all experimental data in Fig. 2 of Sheet et. al. do show the same energy scale (i.e., provide energy-resolved spectroscopic info) no matter whether the dips occur or not. This fact was already clearly stated in our first response letter (Sec. I).

Furthermore, we would like to point out that Sheet et al. themselves concluded in the abstract of their paper that "*such dips are caused by the contact not being in the ballistic limit*", and in fact they explained the occurrence of the dips mostly in the diffusive regime (which is a SPECTROSCOPIC regime) with mixed contribution from Sharvin (ballistic) resistance $R_s$ and the Maxwell (inelastic scattering) resistance $R_M$. Sheet et al. further noted [page 5 of PHYS. REV. B 69 134507 (2004)]: "*In principle, one could also think of an opposite situation where the contact is in the ballistic limit but where the superconductor reaches Ic at voltage values smaller than or of the order of Δ/e. Since the features unambiguously* *associated with Andreev reflection occur in the voltage range 2Δ/e, one would get a spectrum with sharp dips and no feature associated with Andreev reflection will appear.*" Therefore, the paper of Sheet et al. clearly stated that the dI/dV dips can occur in both diffusive and ballistic regimes (all spectroscopic regimes giving accurate gap energy info), which contradicts the claim that "*these (our) experiments do not provide any spectroscopic information*" due to the occurrence of dI/dV dips.

2. One may states: "*Furthermore, the authors make a wrong statement: "conductance dips occur when the critical current Ic of the corresponding section of the SC microcrystal is reached, and have nothing to do with the transport regime"—one should note that the critical current will play a role only when the bulk resistivity of the material contributes appreciably to the point-contact resistance and that does not happen in the ballistic (Sharvin) regime of transport. The Sharvin resistance is independent of the bulk resistivity. If one observes the dips associated with critical current, one is in the thermal regime where the Maxwell's resistance dominates and the bulk resistivity (and the critical current) has a role to play. Therefore, I believe that the authors have drawn conclusions without fully understanding the regimes of their transport experiments.*"

The above words that we have made a "wrong statement" reflect the fact that he/she missed the subtle point in understanding the role of the critical current in the dI/dV dips, which has been clarified previously by diagnostic experiments [see Fig. 3 and the discussion in page 4 of J. Phys.: Condens. Matter 24, 455703 (2012)]. The argument that, "*one should note that the critical current will play a role only when the bulk resistivity of the material contributes appreciably to the point-contact resistance and that does not happen in the ballistic (Sharvin) regime of transport*", is not true. One has to consider the size of the superconducting grain, i.e., the corresponding section of the SC microcrystal in our experiments (or an unintentionally formed superconducting grain in traditional needle-anvil point-contact AR method). As explained previously in details [J. Phys.: Condens. Matter 24, 455703 (2012)], when the superconducting grain reaches normal state at the critical current Ic, a dI/dV dip (solely from the contribution of the N-S interface, without contribution from the bulk resistivity at all) occurs naturally since an inflection point of the I-V curve must exist in order to connect two I–V sections (before and after Ic) smoothly, even if the N-S interface itself is ballistic. This is a simple mathematical fact in calculus, and supported by experimental data showing the concurrent occurrence of the dI/dV dip and the critical current Ic of the corresponding SC microcrystal (grain) [J. Phys.: Condens. Matter 24, 455703 (2012)].

Also, as already discussed in point #1 above, the paper of Sheet et al clearly shows that it is simply wrong in further asserting, "*If one observes the dips associated with critical current, one is in the thermal regime where the Maxwell's resistance dominates and the bulk resistivity (and the critical current) has a role to play.*"

3. One may states: "*In addition, to address most of my objections the authors used the phrase "standard practice"! They failed to appreciate that when the Gamma value is comparable to the gap value itself, the fitting makes no sense as in that case the lifetime of the quasiparticles will be so small that superconductivity itself will be almost* *destroyed. A back of the envelope calculation based on the uncertainty principle should be enough to establish this."*

One cannot ignore the fact that comparable fitting parameters have been reported in literatures to explain experimentally observed broadening, as we pointed out in page 8 of the first response letter. The key point is that the quasi-particle lifetime only partly contributes to the broadening. Therefore, a relatively large broadening parameter does NOT necessarily mean a small quasiparticle lifetime. As we already stated in the first response letter (Sec. I), "*the physical origin of the broadening parameter $\Gamma$ includes different sources [see: Supercond. Sci. & Tech. 23, 043001, 2010 and Ref. 36 therein], such as intrinsic quasiparticle lifetime, inelastic scattering near the N-S interface (e.g., due to surface degradation or contaminations), and even a distribution of gap values (e.g. in anisotropic superconductors).*" Therefore, the above physical reasoning is not justified.

4. One may concludes: "*Therefore, I believe that most of the spectra presented in the manuscript do not provide any spectroscopic information and the fitting procedure that the authors have followed is invalid. No conclusion can be drawn about the physics in topological superconductors from the presented set of experiments.*"

In contrast, the above scientific issues have been clearly addressed in our previous response letter considering: (1) the experimental data and physical analysis in existing literatures does NOT support an assertion that our AR data "do not provide any spectroscopic information" due to the occurrence of dI/dV dips; and (2) comparable fitting parameters have been reported in literatures to explain experimentally observed broadening and the physical reasoning based on sole contribution from quasiparticle life time is incomplete.

References

1. H.B. Peng, D. De, B. Lv, F. Wei, C.W Chu, arXiv:1301.1030
2. S. Sasaki, M. Kriener, K. Segawa, K. Yada, Y. Tanaka, M. Sato, Y. Ando, Phys. Rev. Lett. **107**, 217001 (2011).
3. S. Sasaki, Z. Ren, A. A. Taskin, K. Segawa, L. Fu, Y. Ando, Phys. Rev. Lett. **109**, 217004 (2012).